\documentclass[final,5p,times,twocolumn]{elsarticle}

\usepackage{amssymb}

\usepackage{lineno}

\begin{document}
\begin{frontmatter}



\title{Indirect Self-Modulation Instability Measurement Concept for the AWAKE Proton Beam}


\author[CERN,TUG]{M. Turner}
\author[CERN]{A. Petrenko}
\author[CERN,CZK]{B. Biskup}
\author[CERN]{S. Burger}
\author[CERN]{E. Gschwendtner}
\author[binp,nsu]{K. V. Lotov}
\author[CERN]{S. Mazzoni}
\author[CERN]{H. Vincke}

\address[CERN]{CERN, Geneva, Switzerland}
\address[binp]{Budker Institute of Nuclear Physics SB RAS, 630090, Novosibirsk, Russia}
\address[nsu]{Novosibirsk State University, 630090, Novosibirsk, Russia}
\address[TUG]{Graz University of Technology, Graz, Austria}
\address[CZK]{Czech Technical University, Prague, Czech Republic}

\begin{abstract}
AWAKE, the Advanced Proton-Driven Plasma Wakefield Acceleration Experiment, is a proof-of-principle R\&D experiment at CERN using a $400\,\textrm{GeV/c}$ proton beam from the CERN SPS (longitudinal beam size $\sigma_z = 12\,\textrm{cm}$) which will be sent into a $10\,\textrm{m}$ long plasma section with a nominal density of $\approx 7\cdot 10^{14}\,\textrm{atoms/cm}^3$ (plasma wavelength $\lambda_p = 1.2\,\textrm{mm}$).
In this paper we show that by measuring the time integrated transverse profile of the proton bunch at two locations downstream of the AWAKE plasma, information about the occurrence of the self-modulation instability (SMI) can be inferred. In particular we show that measuring defocused protons with an angle of $1\,\textrm{mrad}$ corresponds to having electric fields in the order of GV/m and fully developed self-modulation of the proton bunch. Additionally, by measuring the defocused beam edge of the self-modulated bunch, information about the growth rate of the instability can be extracted. If hosing instability occurs, it could be detected by measuring a non-uniform defocused beam shape with changing radius. Using a $1\,\textrm{mm}$ thick Chromox scintillation screen for imaging of the self-modulated proton bunch, an edge resolution of $0.6\,\textrm{mm}$ and hence a SMI saturation point resolution of $1.2\,\textrm{m}$ can be achieved.

\end{abstract}


\end{frontmatter}


\section{Introduction}
The AWAKE \cite{AWAKE} (Advanced Proton-Driven Plasma Wakefield Acceleration Experiment) experiment currently being built at CERN intends to use a $400\,\textrm{GeV/c}$ proton bunch extracted from the SPS to accelerate electrons in plasma wakefields with amplitudes up to the GV/m level.\\
In order to drive plasma wakefields efficiently, the length of the drive bunch has to be on the order of the plasma wavelength. Since this is not the case for this proton beam with $\sigma_z = 12\,\textrm{cm}$, the experiment relies on the self-modulation instability (SMI)\,\cite{SMI}, which modulates the proton driver at the plasma wavelength in the first few meters of plasma.
AWAKE is going to use a $10\,\textrm{m}$ long rubidium vapor cell \cite{PLASMA} filled with a density of $7\cdot10^{14}$ atoms/$\textrm{cm}^3$, which corresponds to a plasma wavelength of $\lambda_p = 2\pi c/\omega_{pe} = 1.2\,\textrm{mm}$, where $c$ is the speed of light and $\omega_{pe} = (4 \pi n_e e^2/m_e)^{1/2}$ is the electron plasma frequency being $= 1.57 \cdot 10^{12}\,\textrm{s}^{-1}$ for AWAKE, $n_e$ is the plasma electron density and $e$ and $m_e$ are the electron charge and mass. A short laser pulse ($100\,\textrm{fs}$, $I = 20\,$ TW/cm$^2$, $a_0 \leq 10^{-2}$) will be used to both ionize the rubidium vapor with a minimum plasma radius of $r_0 \approx 1.5\,\textrm{mm}$, and to seed the SMI of the proton bunch through a sudden onset of the plasma density. The proton drive bunch contains $3\cdot10^{11}$ particles.\\ 
The SMI is a transverse two-stream instability that modulates a long drive bunch into micro-bunches with the plasma frequency \cite{SMIPLASMA}. The initial wakefield is created by the sudden onset of the plasma density created by laser pulse. Because of the focusing and defocusing transverse fields created by the wakefields, protons in the focusing phase of the wakefields exit the plasma close to the plasma center and appear as a narrow core on transverse profiles downstream from plasma. Protons in the defocusing phase of the wakefields appear as a halo surrounding the core.
According to plasma simulations \cite{TRAP} these fields can be on the order of GV/m resulting in a defocusing angle on the order of $1\,\textrm{mrad}$ for the $400\,\textrm{GeV/c}$ protons.\\
The first phase of the AWAKE experiment will start in autumn 2016 and the goal is to study the SMI.\\
In this paper we show that by measuring the transverse proton bunch profiles at two locations downstream from the plasma, it can be proven that SMI developed successfully in the plasma and hence strong transverse fields were created. Furthermore, some information about the growth-rate of the instability can be obtained by tracking particles back and measuring where the protons experienced transverse kicks. By looking at the beam image itself it can be seen if hosing instability developed. \\
In addition, technical parameters about the screen, the interaction of the screen with the proton beam and the expected resolution for the angle measurement are presented.

\section{The indirect SMI measurement principle}
\begin{figure}[htb!]
		\includegraphics[width = 0.5\textwidth]{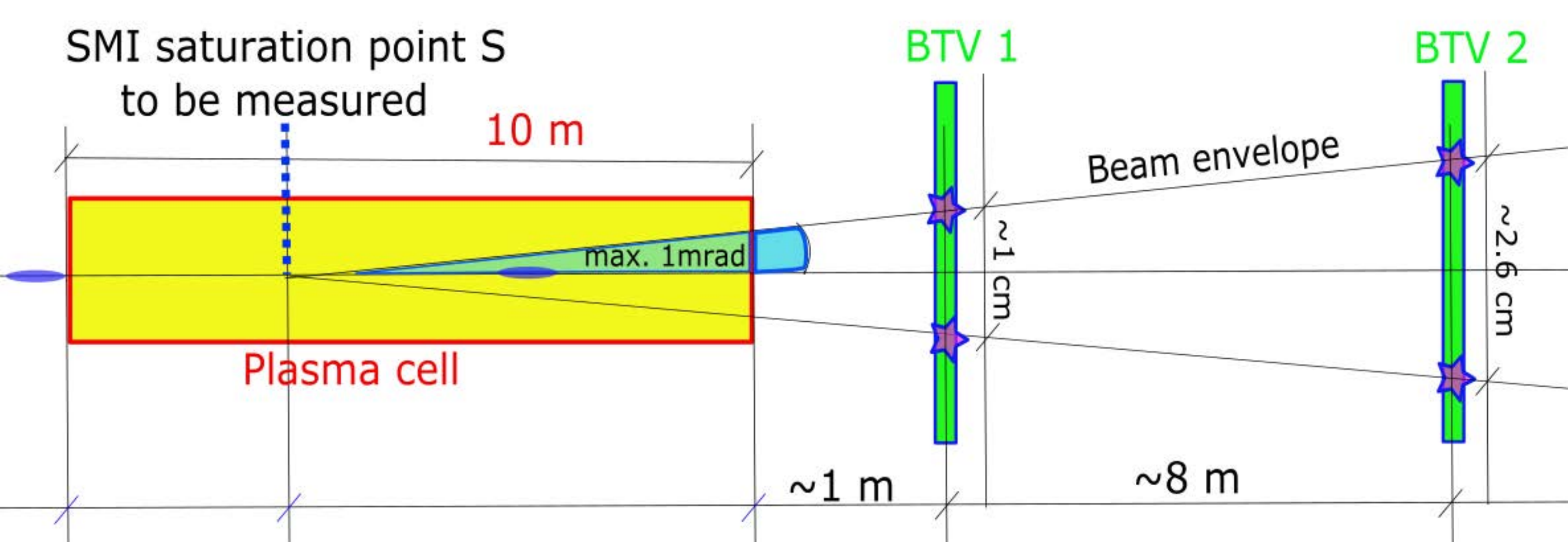}
		\caption{Schematic drawing of the measurement setup in AWAKE. Two measurement screens (BTV 1,2) are inserted downstream the plasma. The distance between the screens is $\approx 8\,\textrm{m}$. S is the SMI saturation point.}
		\label{fig:layout}
\end{figure}

The basic idea of the indirect SMI measurement is to insert two beam-imaging screens (BTVs) downstream of the plasma in order to measure the transverse bunch shape and the beam size at two locations $\approx 8\,\textrm{m}$ apart (see Fig. \ref{fig:layout}).
The following information can be extracted from the measurement.\\
\subsection{Estimation of transverse wakefield strength}
Defocused protons experience transverse kicks with an angle $\theta$ in the plasma. The order of magnitude of the defocusing angle can be estimated as follows. The defocusing angle $\theta$ is defined as:
\begin{equation}
\label{eq:1}
\theta = \frac{\Delta p_r}{p_z}
\end{equation}
where $\Delta p_r$ is the change of radial momentum and $p_z$ the longitudinal momentum. Under the influence of an electric field $E_r$ protons move according to Newton's equation:
\begin{equation}
\label{eq:2}
\frac{\Delta p_r}{\Delta t} \approx e E_r \Rightarrow \Delta p_r \approx e E_r \Delta t
\end{equation} 
with $\Delta t$ being the time that the particle takes to travel that transverse distance $\Delta r$. In this particular case $\Delta p_r$, the change of radial momentum, is due to transverse electric fields $E_r$, which can reach a fraction of the wave-breaking field $E_0$, the maximum electric field that can be generated at a given plasma density:
\begin{equation}
 E_0 = \frac{m_e c^2}{e (c/\omega_{pe})}.
\label{eq:22}
\end{equation}
From eq. (\ref{eq:22}) one can see that $E_0$ is the electric field which accelerates an electron to the energy of $m_e c^2$ over a distance of $c/\omega_{pe}$. 
Assuming that the transverse electric field reaches a significant fraction of the wave-breaking field, let us say $E_r = E_0/2$ and using $\Delta r/\Delta z \sim \Delta p_r/p_z$, $\Delta t$ can be estimated as:
\begin{equation}
\Delta t \sim \sqrt{\frac{p_z}{e E_r}\frac{\Delta r}{c}}
\label{eq:3}
\end{equation}
with $\Delta r$ being the plasma channel width of $\sim 2\,c/\omega_{pe}$.
Combining eq. (\ref{eq:3}) with eq. (\ref{eq:1}) and eq. (\ref{eq:2}) and $p_z = \gamma m_p c$ with $m_p$ being the proton mass and assuming that the proton bunch moves with the speed of light, the following expression for the defocusing angle $\theta$ can be obtained:
\begin{equation}
\theta = \frac{\Delta p_r}{p_z} \sim \sqrt{\frac{m_e}{\gamma m_p}}.
\end{equation}
Using the AWAKE parameters of a proton beam with $400\,\textrm{GeV/c}$ corresponding to $\gamma = 427$, $\theta$ is estimated to be $\approx 1\, \textrm{mrad}$. For example for LHC parameters ($\gamma = 7478$) $\theta$ would be around $0.3\,\textrm{mrad}$.\\
Note that according to this estimate the maximum angle of the defocused protons does not depend on the plasma density, provided that the wakefield amplitude reaches half of $E_0$ also independently of the plasma density.
Plasma simulation results obtained with LCODE \cite{LCODE1,LCODE2} show that protons experience angular kicks of about $1\,\textrm{mrad}$ in AWAKE.\\ 
This means that observing diverging angles in the order of $1\,\textrm{mrad}$, as opposed to $\approx 0.01\,\textrm{mrad}$ without plasma, from the two beam images is a proof that high amplitude plasma wakefields are excited by SMI.

\subsection{Understanding the SMI growth rate}
The saturation point of the SMI (as illustrated in Fig. \ref{fig:layout}), which is the point of the maximum wakefield amplitude, can be estimated by connecting the beam edges on the two screens and tracking back to the crossing point inside the plasma cell. This assumes that the protons will get most of the kick close to the point of maximum wakefield amplitude. According to simulations the saturation point is around $3000\,\lambda_p$ which is around $4\,\textrm{m}$ for a plasma density of $7\cdot10^{14}$ atoms/cm$^3$. The worst case resolution of the saturation point can be calculated by using the simulated resolution of the defocused beam edge of $0.6\,\textrm{mm}$ to $\Delta S = 1.2\,\textrm{m}$, which was obtained as will be described in section \ref{resolution}. This error can be improved by taking more than one measurement of the defocused beam radius or by combining independent measurements. 

\subsection{Detecting hosing instability}
More information can be extracted from the measurement by looking at the beam edge shape on the screen. A round defocused beam-edge shape is expected if SMI dominates, because the SMI is cylindrically symmetric. If the hosing instability \cite{HOSING,HOSING2} developes, the defocused beam-edge shape is expected to be smaller and changing in radius, because wakefields are weaker and non axis symmetric. The wakefields are weaker because the number of micro-bunches driving the wakefield resonantly is less than when only SMI developed.

\section{Design of the imaging screens}
\label{sec:bgandscat}
The proposed screen material for this experiment is the scintillator material Chromox ($Al_2O_3:CrO_2$). It has a light yield of $10^4$ photons per MeV of deposited energy \cite{LY} that is suitable for detecting the radius of the defocused beam. The chosen screen thickness is $1\,\textrm{mm}$ (explained in more detail in section \ref{sec:scat}). The material is pink, thick and opaque which means that photons that will be detected are emitted from a surface layer.
\subsection{Layout}
\label{sec:layout}
\begin{figure}[htb!]
		\includegraphics[width = 0.45\textwidth]{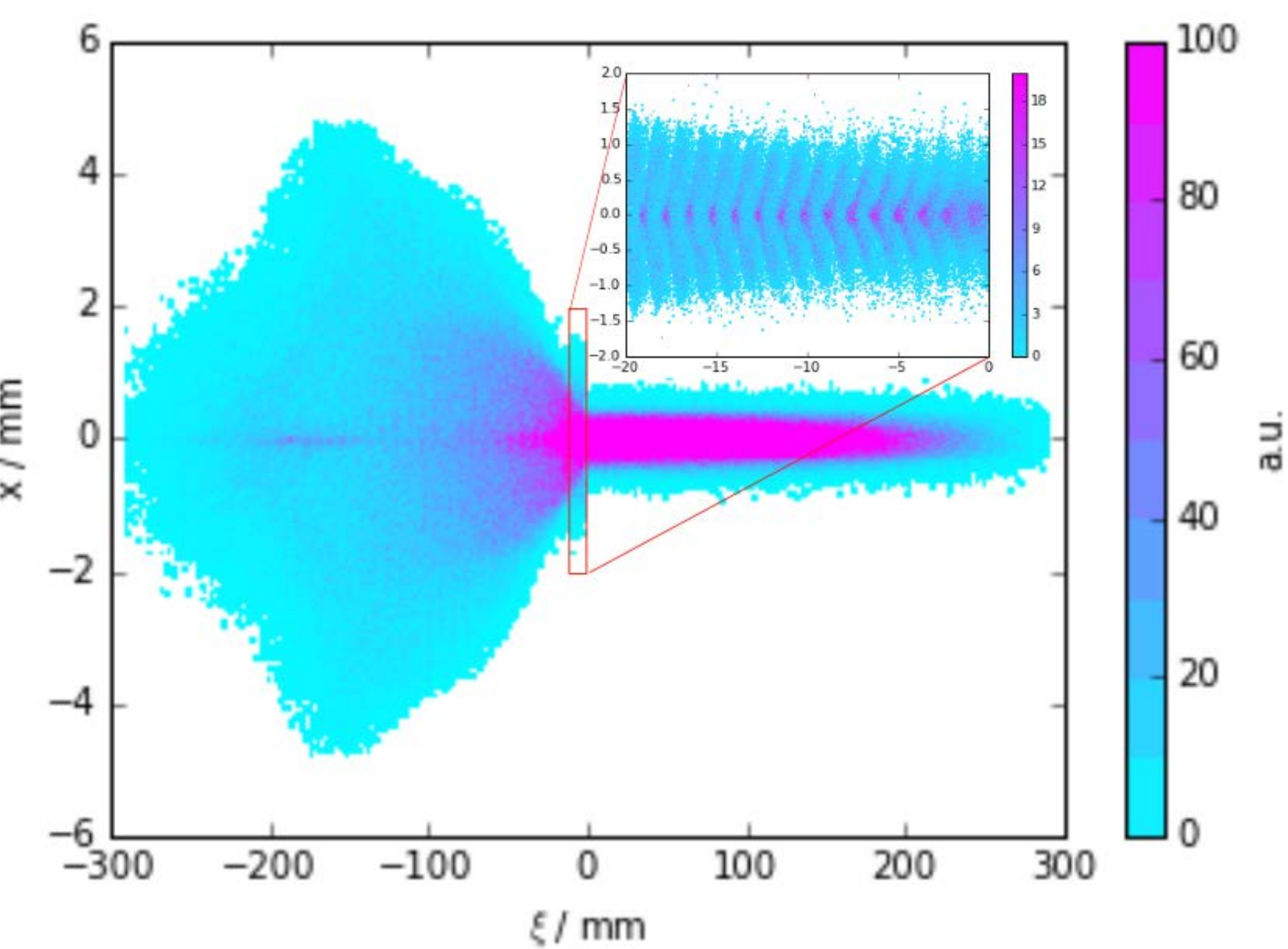}
		\caption{2d density plot of the proton beam coming from plasma simulations with LCODE used as an input for the FLUKA simulations. The position of the laser pulse at $\xi = 0$ separates the bunch into having a gaussian transverse shape in the front and self-modulated microbunches surrounded by defocused particles in the second half. The zoomed region shows the micro bunched structure of the beam core after the laser pulse. The bunch propagates towards the right.}
		\label{fig:input}
\end{figure}

Simulations of the interactions of the proton bunch with the scintillator screens were performed with FLUKA \cite{FLUKA}. FLUKA is a bench-marked Monte-Carlo code that simulates particle interactions with matter. The simulation geometry comprises the laser dump ($1\,\textrm{mm}$ of aluminum), used to stop the laser pulse $50\,\textrm{cm}$ upstream the first screen, the two imaging screens (each $1\,\textrm{mm}$ of alumina) and a vacuum pipe (iron) with a diameter of $80\,\textrm{mm}$. The simulation input beam is shown in Fig. \ref{fig:input} and was obtained from plasma simulations performed with LCODE, using the AWAKE baseline parameters. The front part of the proton bunch is transversely gaussian, but the second half of the bunch is self-modulated so that the bunch has a micro-bunched core surrounded by defocused particles \cite{TRAP}.

\subsection{Background studies}
\label{sec:scat}
Every material that the proton bunch has to traverse will scatter protons and produce secondary particles and hence complicates the beam shape measurement, since secondary particles with distributions not related to SMI also produce light in the screen. Analytical calculations as well as FLUKA simulations have been performed in order to estimate the significance of the scattered protons and secondary particles produced. 
The number of scattered protons in the interesting angle range from $10^{-2}$ to $10^{1}\,\textrm{mrad}$ is estimated using the analytical calculations based on Coulomb scattering in the order of 1 scattering per 100 protons. This angle range was selected because protons experiencing larger than $\approx 10^{1}\,\textrm{mrad}$ will be out of the screen. Kicks smaller than $\approx 10^{-2}\,\textrm{mrad}$ cannot be resolved because the transverse distance they travel is smaller than the resolution of the screen.
FLUKA simulations show that the main background contributions are coming from secondary photons and pions created in the screen. The photon production ($E_{\gamma} \sim 0.001-400\,\textrm{GeV}$) is significant but according to \cite{PHOINT} only $\approx 1\,\%$ of these photons will interact with the screen.\\
The radiation length in aluminum is $\approx 40\,\textrm{cm}$. The number of scattered protons and secondary particles increases linearly with the material thickness. Since the changes of screen thicknesses are in the sub-mm range, the change of number of scattered protons and secondary particles is small. Also the number of scintillation photons produced depends only on the ionization created in the surface layer. Consequently the impact of the screen thickness on the number of photons detected and thus on the bunch shape is expected to be negligible. This was simulated and confirmed by FLUKA simulations studying $0.25$, $0.5$, $1$ and $1.5\,\textrm{mm}$ thick screens, thus the standard screen thickness of $1\,\textrm{mm}$ was selected.
\subsection{Deposited Energy in the screens}
\begin{figure}[htb!]
		\vfill{a)}\\
		\includegraphics[width = 0.45\textwidth]{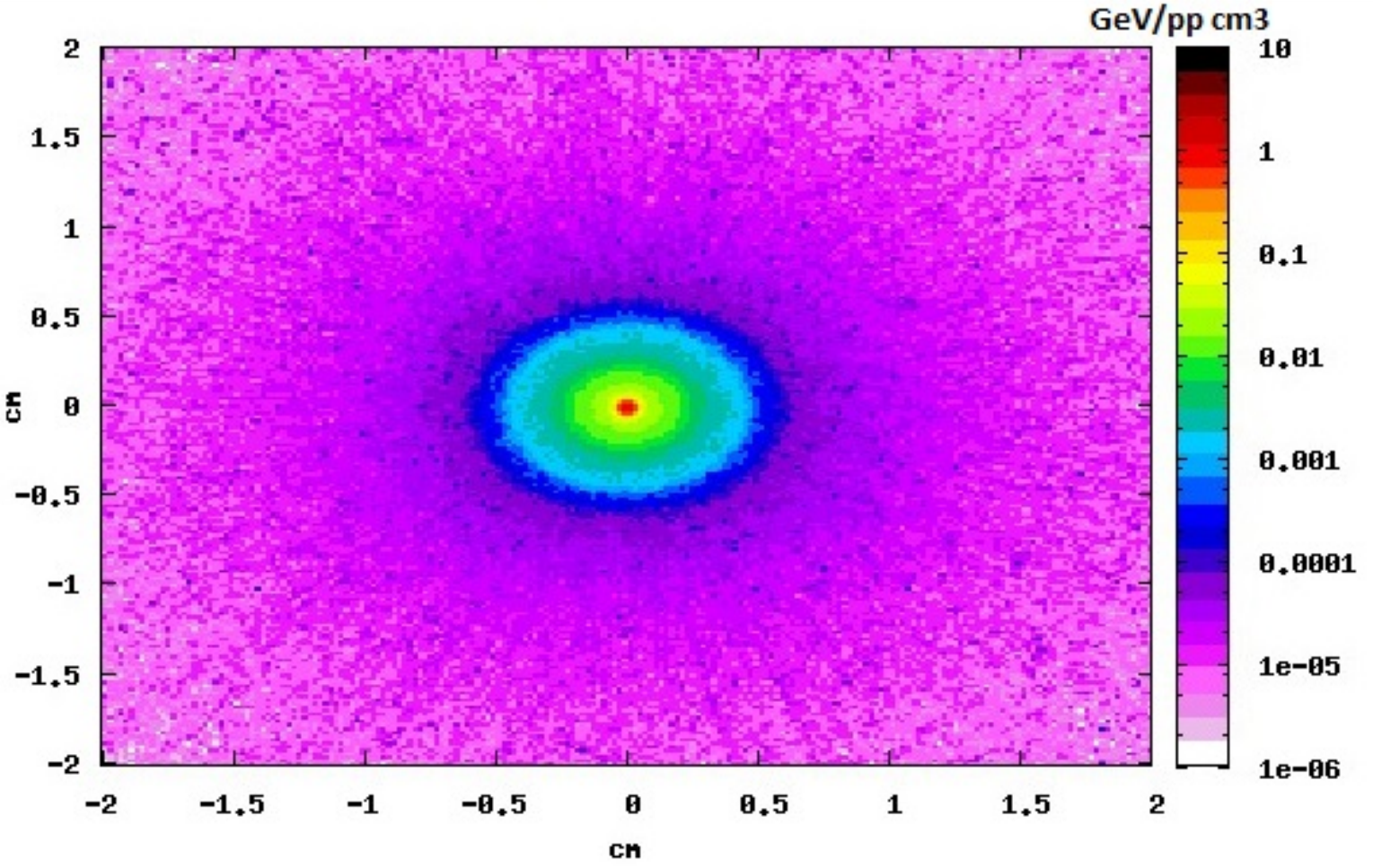}
		\vfill{b)}\\
		\includegraphics[width = 0.45\textwidth]{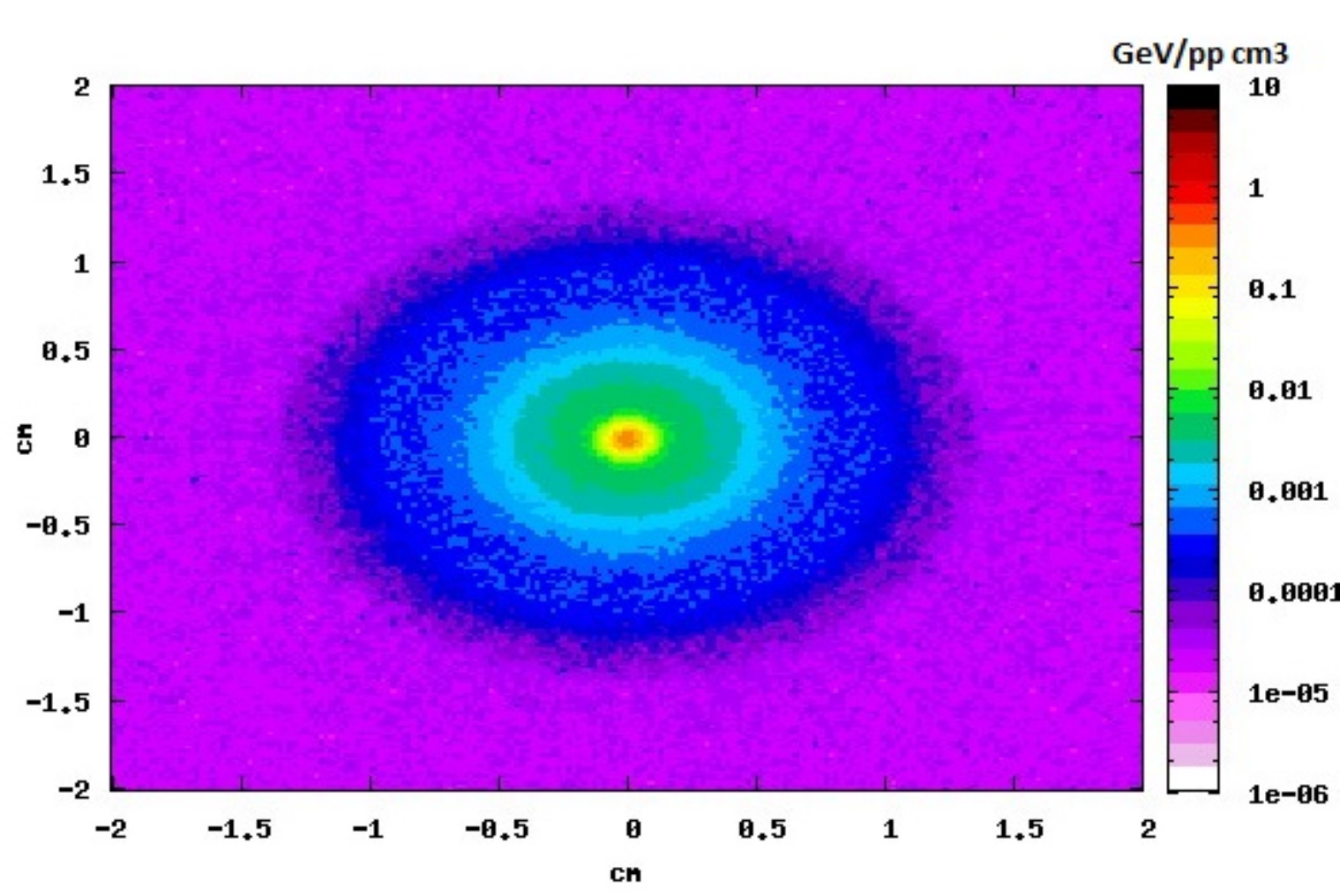}
		\caption{Energy deposition of the self-modulated $400\,\textrm{GeV}$ proton beam in a) the first and b) the second $1\,\textrm{mm}$ thick Chromox screen. These results were obtained from FLUKA simulations using the input beam shown in Fig. \ref{fig:input}.}
		\label{fig:energydep}
\end{figure}

The energy deposit of a $400\,\textrm{GeV/c}$ proton beam in an $1\,\textrm{mm}$ thick Alumina ($Al_2O_3$, not doped for simplicity) screen was simulated with FLUKA, using the same geometry layout as described in section \ref{sec:layout}. The energy deposition in the screen (see Fig. \ref{fig:energydep}) is important because the number of optical photons produced in a scintillator is proportional to the total deposited energy. Particles passing the screen ionize the screen material, and some part of this excess energy radiates as optical photons. Furthermore the screen temperature increases with each proton bunch and it has to be insured that the screen does not reach the $1500\,^\circ \textrm{C}$ operation limit.
 The temperature rise in the screen was calculated:
\begin{equation}
E_{dep} = \int_{T_1}^{T_2} \rho c(T) dT \Rightarrow \Delta T = \frac{E_{dep}}{\rho  c} \sim 20\,K
\end{equation}
with $E_{dep}$ being the energy deposit per unit volume, $\rho$ the density of the screen material, and $c(T)$ the heat capacity of the material as a function of temperature. In our case the heat capacity was assumed to be constant, because of the minor temperature rise that was obtained.
The maximum deposited energy from Fig. \ref{fig:energydep}: $E_{dep} =1\,\textrm{MeV/mm}^3$ per primary particle per screen. 
The screen density is $2.7\cdot 10^{-3}\,\textrm{g/mm}^3$ and the heat capacity is assumed constant with $c = 0.88\,\textrm{J/gK}$. 
Chromox is a material commonly used in the CERN accelerators and has been operated without any issues so far.

\subsection{Screen resolution}
\label{resolution}
The intrinsic screen resolution coming from the CCD readout chip is $25\,\mu \textrm{m}$ for a $1\,\textrm{cm}^2$ chip with 400x400 pixel. Combining this intrinsic resolution with the blurring from ionization, the photon production and the optics, the achievable resolution is around $100\,\mu\textrm{m}$. Optimally with the measurement one can detect the edge location of the defocused beam. The intensity of the beam core is approximately five orders of magnitude higher than the defocused beam and hence the defocused beam edge is challenging to detect. Optimization studies are ongoing.
Standard CCD cameras can detect a dynamic range of two to three orders of magnitude. In order to keep the sensitivity to measure the edges, one possibility would be to cut a hole into the screen to exclude the bright center, another possibility is to make a combined screen with a material of low light yield in the center and a material with high light yield around.
Nevertheless proton scattering and secondary particles arriving on the screen smear the edge and make the detection difficult.
\begin{figure}[htb!]
\centering
\hspace{-0.5cm}
		\includegraphics[width = 0.5\textwidth]{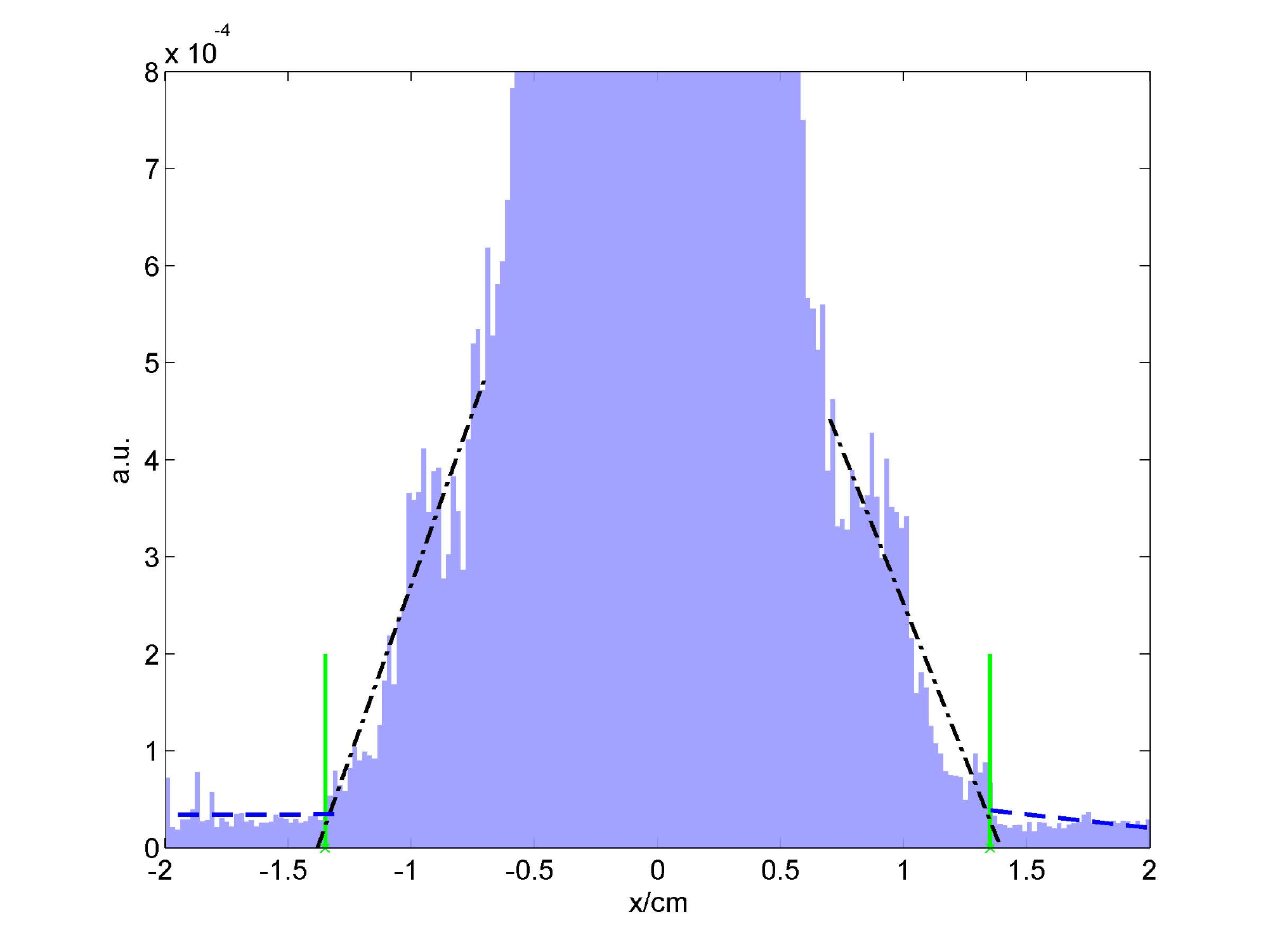}
		\caption{Illustration of the linear fitting procedure of the obtained signal. The blue dashed lines represent the linear background fits. The black dashed-dotted lines are the linear fits of the linear signal region and the green solid line marks the x-position of the undisturbed defocused beam edge.}
		\label{fig:fit}
\end{figure}
The simulated proton beam density has a linear dependence of radius close to the sharp beam edge. This dependency can be used to detect the beam edge if linear fits are applied to the outer signal region and the background of the simulated optical photon signal (see Fig. \ref{fig:fit}). Taking the crossing point of these linear fits and comparing them to the defocused proton beam edge (which is known by tracking the undisturbed defocused beam from the simulations) gives an error of $0.6\,\textrm{mm}$. This number was obtained by taking the standard deviation of 16 fitted crossing points with the real edge coming from simulations.  
\section{Conclusions}
Implementing two scintillating screens downstream of the plasma in the AWAKE experiment opens the possibility to prove that the proton bunch experiences SMI by detecting defocused protons on the images. The expected angular divergence for the AWAKE experiment is at the order of $1\,\textrm{mrad}$. Information about the growth rate of SMI can be extracted from the beam images by extrapolating the saturation point of the SMI. Additionally if hosing instability occurred, it could be detected by measuring a non-uniform defocused beam shape with changing radius. By applying a linear fit to the the simulated beam signals, a defocused beam edge resolution of $0.6\,\textrm{mm}$ is achieved. This corresponds to a saturation point resolution of $1.2\,\textrm{m}$. The preliminary measurement design consists of two $1\,\textrm{mm}$ thick Chromox screens with either a hole or a low light yield material at the bright beam center. 


\section{Acknowledgment}
The authors would like to thank P. Muggli for his very useful corrections. The contribution of K.Lotov to this work is supported by The Russian Science Foundation (grant No. 14-12-00043).
\section{References}



\end{document}